\documentclass[fleqn,12pt,twoside]{article}
\usepackage{espcrc1}

\usepackage{graphicx}
\usepackage[figuresright]{rotating}

\newcommand{\lap}{\mathrel{ \rlap{\raise.5ex\hbox{$<$}}
                    {\lower.5ex\hbox{$\sim$}}  } }

\newcommand{\fo}{$^{18}$F}
\newcommand{\fn}{$^{19}$F}
\newcommand{\nen}{$^{19}$Ne}

\newcommand{\pa}{$^{18}$F(p,$\alpha)^{15}$O}
\newcommand{\dpa}{D($^{18}$F,p$\alpha)^{15}$N}

\newcommand{\zaa}{{\it Astron. Astrophys.}}

\newcommand{\zapj}{{\it Astrophys. J.}}

\newcommand{\zpr}{{\it Phys.~Rev.}}
\newcommand{\zprl}{{\it Phys.~Rev.~Lett.}}
\newcommand{\znim}{{\it Nucl.~Inst.~and~Meth.}}

\newcommand{\zmnras}{{\it MNRAS}}

\newcommand{\AmS}{{\protect\the\textfont2
  A\kern-.1667em\lower.5ex\hbox{M}\kern-.125emS}}

\title{The \pa\ reaction rate for application to nova $\gamma$-ray emission}

\author{N.~de~S\'er\'eville\address[CSNSM]{CSNSM, CNRS/IN2P3/UPS, B\^at.~104, 91405 Orsay Campus, France}\thanks{Presently at IEEC, Edifici Nexus 201, C/ Gran Capita 2-4, 08034 Barcelona, Spain},
        E.~Berthoumieux\address[CEA]{CEA, DAPNIA/SPhN, F-91191 Gif/Yvette 
	Cedex, France}
	and
        A.~Coc\addressmark[CSNSM]
        }

\begin{document}

% typeset front matter
\maketitle

\begin{abstract}
The \pa\ reaction is recognized as one of the most important reaction for
nova gamma--ray astronomy as it governs the early $\leq$ 511~keV
emission. However, its rate remains largely uncertain at nova
temperatures due to unknown low--energy resonance strengths. 
We report here on our last results concerning the study of the \dpa\ reaction, 
as well as on the determination of the \pa\ reaction rate using the R-matrix 
theory. Remaining uncertainties are discussed.
\end{abstract}

\section{INTRODUCTION}

Gamma--ray emission from classical novae is dominated, during the
first hours, by positron annihilation following the beta
decay of \fo\cite{Gom98,Her99,Coc00}. However, even though it has been 
the object of many recent experiments\cite{Gra00,Bar01,Bar02} the rate 
of its main mode of destruction, through the \pa\ reaction, remains 
highly uncertain. This was mainly due to the unknown proton widths of 
the first three \nen\ levels above the proton emission threshold ($E_x$, 
$J^\pi$ = 6.419~MeV, 3/2$^+$; 6.437~MeV, 1/2$^-$ and 6.449~MeV, 3/2$^+$).
The tails of the corresponding resonances (at respectively, $E_R$ =
8, 26 and 38~keV) can dominate the astrophysical S-factor in
the relevant energy range\cite{Coc00}. As a consequence of these
nuclear uncertainties, the \fo\ production in nova and the early
gamma--ray emission was uncertain by a factor of $\approx$300\cite{Coc00}.
Since a direct measurement of the relevant resonance strengths 
is impossible due to the very low Coulomb barrier penetrability, we
used an indirect method aiming at determining the one--nucleon 
spectroscopic factors in the analog levels of the mirror nucleus (\fn) 
by the neutron transfer reaction D($^{18}$F,p)$^{19}$F\cite{Ser03}. 
Recently, the same reaction has been studied at higher energy\cite{Koz04}.
We present here additional information on the extraction of spectroscopic 
factors as well as new \pa\ reaction rates.

\section{EXPERIMENT AND ANALYSIS}

\begin{figure}[ht]
  \centering
  \includegraphics[width=15cm]{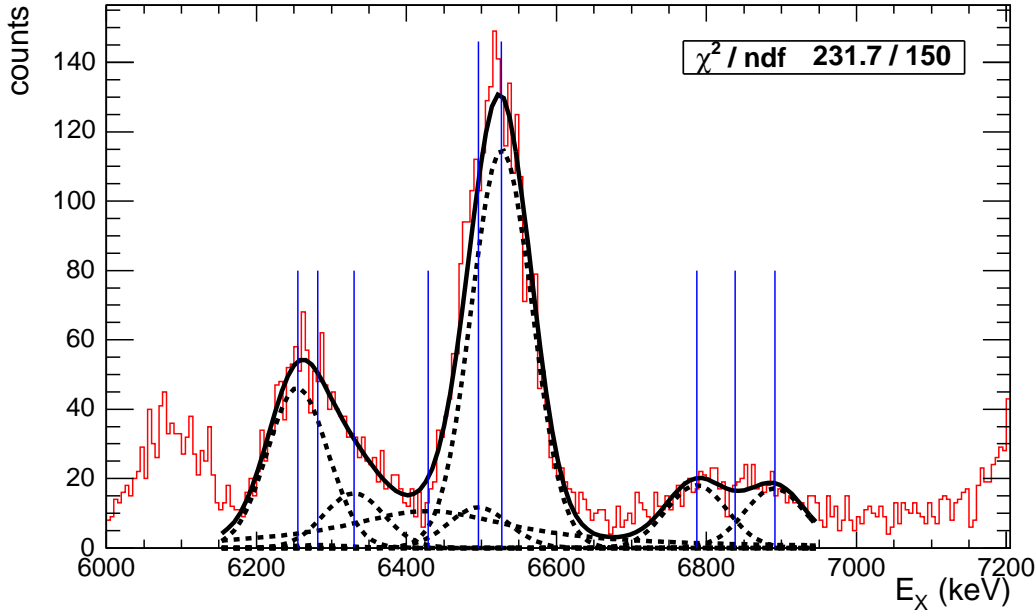}
  \caption{\fn\ excitation energy calibration spectrum with a global fit 
  (see text). The $E_X = 6.527$~MeV level is the most populated.}
  \label{f:spect}
\end{figure}

We refer to de~S\'er\'eville et al.\cite{Ser03} for experimental details
and present in Figure~\ref{f:spect} the reconstructed excitation energy 
spectrum. The well populated \fn\ levels at $E_X = 2.780$ and 5.106~MeV 
observed in single events as well as the $E_X = 7.262 + 7.368$~MeV group 
observed in coincidence events were used to make an internal calibration. 
This procedure resulted in an uncertainty on the excitation energy calibration 
of $\approx 3$~keV in the region of interest. During this analysis a careful 
study of the systematic errors has been done. To extract the relative 
contribution of the two 3/2$^+$ levels, a simultaneous fit of the 6.5~MeV 
group ($E_X$ = 6.497 + 6.528~MeV), the 6.25~MeV group ($E_X$ = 6.255 + 6.282 
+ 6.330~MeV) and the 6.9~MeV group ($E_X$ = 6.787 + 6.838 + 6.891~MeV) was 
performed. The background is described by a lorentzian of width $\Gamma$ = 
280~keV corresponding to the $1/2^-$ $E_X =$~6.429~MeV. The result favors the 
dominant contribution of one single level ($E_X = 6.527$~MeV) while another 
more recent study~\cite{Koz04} favors the other one. In any case, the nuclear 
structure of these two 3/2$^+$ levels seem to be very different according to 
the results of an inelastic electron scattering measurement on \fn~\cite{Bro85}.
From the angular information of the 6.5~MeV peak, we have obtained the 
angular distribution that we have analyzed performing
a finite range DWBA analysis (FRESCO\cite{Tho88} code), including a compound 
nucleus component. The extracted spectroscopic factor is $C^2S = 0.17$ (0.21 
when neglecting the compound nucleus contribution\cite{Ser03}) and is weakly 
dependent of the optical potential parameters. It is to be noted that
our experiment has been recently repeated at higher energy\cite{Koz04} and
that a slightly lower spectroscopic factor is found ($C^2S = 0.12$).

\section{REMAINING UNCERTAINTIES AND REACTION RATE}

Before deriving a new \pa\ reaction rate and because the previous analysis 
showed that the 3/2$^+$ levels of interest {\it cannot} be neglected, it 
seems important to focus on the remaining uncertainties.
First, the spectroscopic factors obtained correspond to \fn\ levels and 
should be transposed to the analog \nen\ levels in order to deduced the 
proton widths used to calculate the reaction rate. Such a common practice 
in nuclear astrophysics leads to an uncertainty of a factor of about two 
on the proton widths (based on a statistical study of analog states in the 
same mass region). Moreover, due to the small separation energy of this 
doublet (only 30~keV), the assignation of analog levels is not very clear and 
in the following we will always deal with two cases: {\it no inversion} where 
the low energy level in \fn\ is the analog of the low energy one in \nen\ 
and {\it inversion} where the low energy level in \fn\ is the analog of the 
high energy one in \nen.
Second, the $\alpha$--widths of the \nen\ levels are unknown and calculated
from the reduced widths of the corresponding analog levels in \fn. Unlike the 
one--nucleon case (see above), the associated uncertainty could be as large
as a factor of 10~\cite{Oli97}. 
Third, the two $3/2^+$ levels of astrophysical interest can interfere with
another $3/2^+$ level at $E_X = 7.076$~MeV and no information about the sign
of interferences is available at present day. In case of destructive 
interference, the astrophysical S-factor decreases drastically in the Gamow 
region at nova temperature.

\begin{figure}[ht]
  \centering
  \includegraphics[width=15cm]{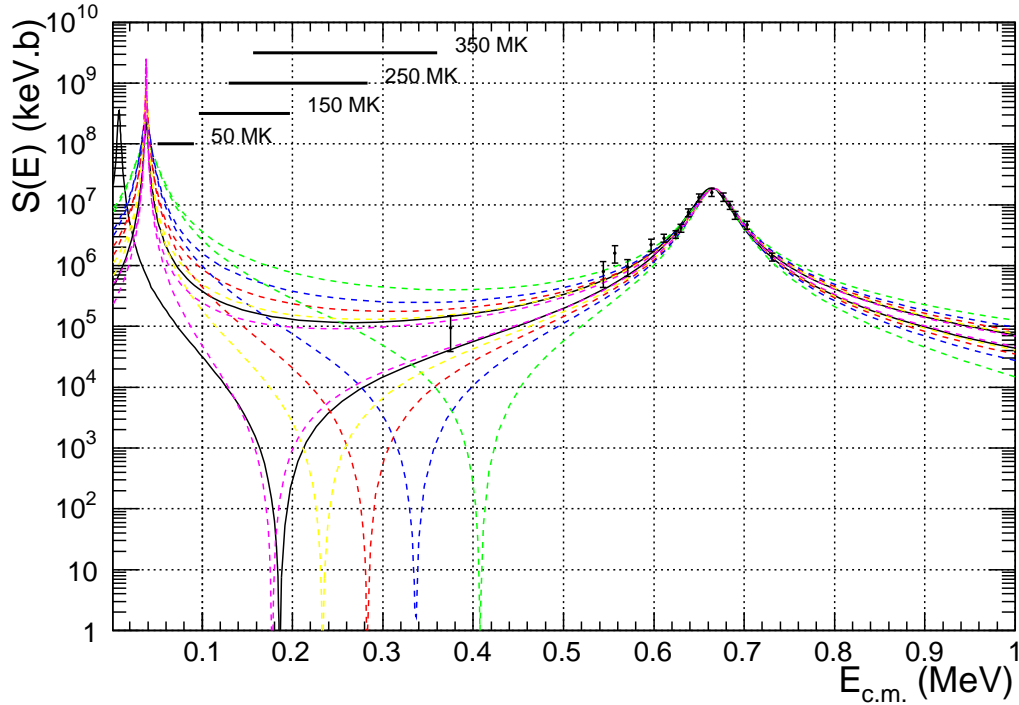}
  \caption{Astrophysical S-factor for constructive and destructive
  interferences between the two \nen\ levels $E_X = 6.449$ and 7.076~MeV
  (dashed lines). Cases for different $\alpha$--width are reported. (The 
  6.449 MeV spectroscopic factor is from our experiment while data points 
  are from Bardayan et al.\protect\cite{Bar02}.). Solid lines represent
  the astrophysical S-factor used for the lower and upper \pa\ reaction
  rate (see text). Gamow windows for temperatures relevant to nova are shown.}
  \label{f:taux}
\end{figure}

Figure~\ref{f:taux} is an example of the influence on the astrophysical 
S-factor of the $\alpha$--width variation when constructive and destructive 
interferences are considered between the two levels $E_X = 6.449$ and 
7.076~MeV (no inversion).
Experimental data from Bardayan et al.\cite{Bar02} are also displayed. These
data points are used to constrain the astrophysical S-factor with the help 
of R-matrix fits (ANARKI~\cite{Ber98} code). The free parameters are the 
$\alpha$--width of the $E_r$ = 8 or 38~keV resonances, the proton width of 
the $E_r$ = 330~keV, the proton and $\alpha$--width and position of the
$E_r$ = 665~keV resonance and the sign of the interferences. The best fit 
is obtained for constructive interferences for the $E_X = 6.449$~MeV level
and is then used for determining the new \pa\ nominal rate. The upper reaction 
rate is given for a constructive interference for the $E_X = 6.419$~MeV 
(inversion) and an $\alpha$--width such as the astrophysical S-factor 
correspond to the upper limit of the error bar of the low--energy data point
from~\cite{Bar02} ($E_{c.m.} = 375$~keV). In the same way, the lower rate is 
given for a destructive interference for the $E_X = 6.449$~MeV (no inversion) 
and an $\alpha$--width such as the astrophysical S-factor correspond to the 
lower limit of the error bar at $E_{c.m.} = 375$~keV.

Since we consider that the inversion of the analog levels is a possibility,
the conclusions are the following. The new nominal rate is within a 
factor of two of the former one~\cite{Coc00}. Furthermore the upper rate 
is reduced and the global uncertainty is reduced but remains large.

\section{ACKNOWLEDGMENTS}
This work is based on experimental results obtained in collaboration with
C.~Angulo, M.~Assun\c{c}\~{a}o, D.~Beaumel, B.~Bouzid, S.~Cherubini, M.~Couder,
P.~Demaret, F.~de~Oliveira~Santos, P.~Figuera, S.~Fortier, M.~Gaelens, 
F.~Hammache, J.~Kiener, D.~Labar, A.~Lefebvre--Schuhl, P.~Leleux, M.~Loiselet, 
A.~Ninane, S.~Ouichaoui, G.~Ryckewaert, N.~Smirnova, V.~Tatischeff and 
J.-P.~Thibaud.

This work has been supported by the European Community-Access to Research
Infrastructure action of the Improving Human Potential Programme, contract
N$^{o}$ HPRI-CT-1999-00110.

\end{document}